\documentclass[pdflatex,sn-vancouver-ay]{sn-jnl}

\usepackage{graphicx}%
\usepackage{multirow}%
\usepackage{amsmath,amssymb,amsfonts}%
\usepackage{amsthm}%
\usepackage{mathrsfs}%
\usepackage[title]{appendix}%
\usepackage{xcolor}%
\usepackage{textcomp}%
\usepackage{manyfoot}%
\usepackage{booktabs}%
\usepackage{algorithm}%
\usepackage{algorithmicx}%
\usepackage{algpseudocode}%
\usepackage{listings}%
\usepackage{bm}
\usepackage{setspace}
\usepackage{lineno}

\begin{document}

\title[Probabilistic SR via a Schrödinger Bridge]{Probabilistic Super-Resolution for Urban Micrometeorology via a Schrödinger Bridge}


\author*[1]{\fnm{Yuki} \sur{Yasuda}}\email{yuki.yasuda@jamstec.go.jp}

\author[2]{\fnm{Ryo} \sur{Onishi}}\email{onishi.ryo@scrc.iir.isct.ac.jp}

\affil*[1]{\orgdiv{Research Institute for Value-Added-Information Generation}, \orgname{Japan Agency for Marine-Earth Science and Technology}, \orgaddress{\street{3173-25 Showa-machi, Kanazawa-ku}, \city{Yokohama}, \postcode{2360001}, \state{Kanagawa}, \country{Japan}}}

\affil[2]{\orgdiv{Supercomputing Research Center, Institute of Integrated Research}, \orgname{Institute of Science Tokyo}, \orgaddress{\street{2-12-1 Ookayama}, \city{Meguro-ku}, \postcode{1528550}, \state{Tokyo}, \country{Japan}}}

\abstract{This study employs a neural network that represents the solution to a Schrödinger bridge problem to perform super-resolution of 2-m temperature in an urban area. Schrödinger bridges generally describe transformations between two data distributions based on diffusion processes. We use a specific Schrödinger-bridge model (SM) that directly transforms low-resolution data into high-resolution data, unlike denoising diffusion probabilistic models (simply, diffusion models; DMs) that generate high-resolution data from Gaussian noise. Low-resolution and high-resolution data were obtained from separate numerical simulations with a physics-based model under common initial and boundary conditions. Compared with a DM, the SM attains comparable accuracy at one-fifth the computational cost, requiring 50 neural-network evaluations per datum for the DM and only 10 for the SM. Furthermore, high-resolution samples generated by the SM exhibit larger variance, implying superior uncertainty quantification relative to the DM. Owing to the reduced computational cost of the SM, our results suggest the feasibility of real-time ensemble micrometeorological prediction using SM-based super-resolution.}




\maketitle

\section{Introduction}\label{sec:introduction}

Deep learning-based super-resolution (SR) has been applied to accelerate numerical weather prediction \citep[e.g.,][]{Onishi+2019, Wang+2021GMD, McGibbon+2024}. In this approach, trained neural networks enhance the resolution of predictions, enabling rapid high-resolution (HR) inference without costly HR numerical integration. Such acceleration has been demonstrated not only for global and mesoscale problems \citep[e.g.,][]{Wang+2021GMD, McGibbon+2024} but also for microscale problems in urban areas \citep{Onishi+2019, Wu+2021, Teufel+2023, Yasuda+Onishi2025a}.

Recently, denoising diffusion probabilistic models \citep[simply, diffusion models (DMs);][]{Ho+2020} have been actively applied to SR in meteorology \citep{Ling+2024, Hess+2025, Mardani+2025, Schmidt+2025, Tomasi+2025}. In DM-based SR, HR samples are generated by repeatedly transforming noise. This sequence of transformations is described by diffusion processes, mathematically formulated as stochastic differential equations (SDEs) with time derivatives parameterized by neural networks \citep{Song+2021a}. Iterative transformations with SDEs yield accurate SR inference \citep[e.g.,][]{Saharia+2023} and allow uncertainty quantification, since the generated samples follow the HR data distribution \citep[e.g.,][]{Mardani+2025}.

SR aims to convert low-resolution (LR) data into HR data, suggesting that DM-based SR becomes more efficient when it starts from LR inputs rather than from noise. The Schrödinger bridge (SB) provides a generalization of DMs for transformations between arbitrary data distributions \citep{Leonard2014}. This problem reduces to estimating SDEs; once an SDE is learned by a neural network, its integration provides the desired transformation \citep[e.g.,][]{DeBortoli+2021, Chen+2022}. In the context of SR, the estimated SDE directly transforms LR data into HR data, leading to greater efficiency than DMs \citep{Liu+2023}.

In meteorology, applications of DMs are rapidly increasing \citep{Ling+2024, Hess+2025, Mardani+2025, Schmidt+2025, Tomasi+2025}, whereas SB-based neural networks remain largely unexplored. In computer vision, where SB methods are advancing, inference accuracy has been widely studied \citep[e.g.,][]{Liu+2023, Wang+2025}, but inference uncertainty has received little attention. Since SB methods also reconstruct data distributions via SDEs, they should allow uncertainty quantification as in DMs. Such evaluation is essential in meteorological problems, which inherently involve uncertainty.

This study applies an SB-based SR model \citep{Chen+2024} to 2-m temperature in an actual urban area. We show that the SB model achieves more efficient probabilistic SR (i.e., ensemble SR inference) than a DM and yields improved ensemble statistics. For clarity, a list of abbreviations is provided in Table A1 of Appendix A.

\section{Probabilistic super-resolution (SR)}\label{sec:probabilistic-SR}

We infer HR data $\bm{x}_\mathrm{HR} \in \mathbb{R}^n$ from LR data $\bm{x}_\mathrm{LR} \in \mathbb{R}^n$ and auxiliary data $\bm{\xi} \in \mathbb{R}^{m\times n}$. All data are interpolated to $n$ grid points at HR using an interpolation method, and $\bm{\xi}$ consists of $m$ variables, such as topographic height. We consider supervised learning, where $\bm{x}_\mathrm{HR}$ serves as the ground truth. HR outputs from neural networks are referred to as HR {\it samples}. Since SR is an inverse problem \citep[e.g.,][]{Park+2003}, the solution is not unique, and multiple HR samples are plausible for a given $\bm{x}_\mathrm{LR}$. This uncertainty is represented by the conditional distribution $p_\mathrm{HR}(\bm{x}\mid \bm{x}_\mathrm{LR}, \bm{\xi})$ \citep[e.g.,][]{Ling+2024}. The SR problem is thus formulated as a probabilistic generative task, with the objective of approximating $p_\mathrm{HR}(\bm{x}\mid \bm{x}_\mathrm{LR}, \bm{\xi})$ using neural networks.

\begin{figure}[ht]
    \centering
    \includegraphics[width=0.9\textwidth]{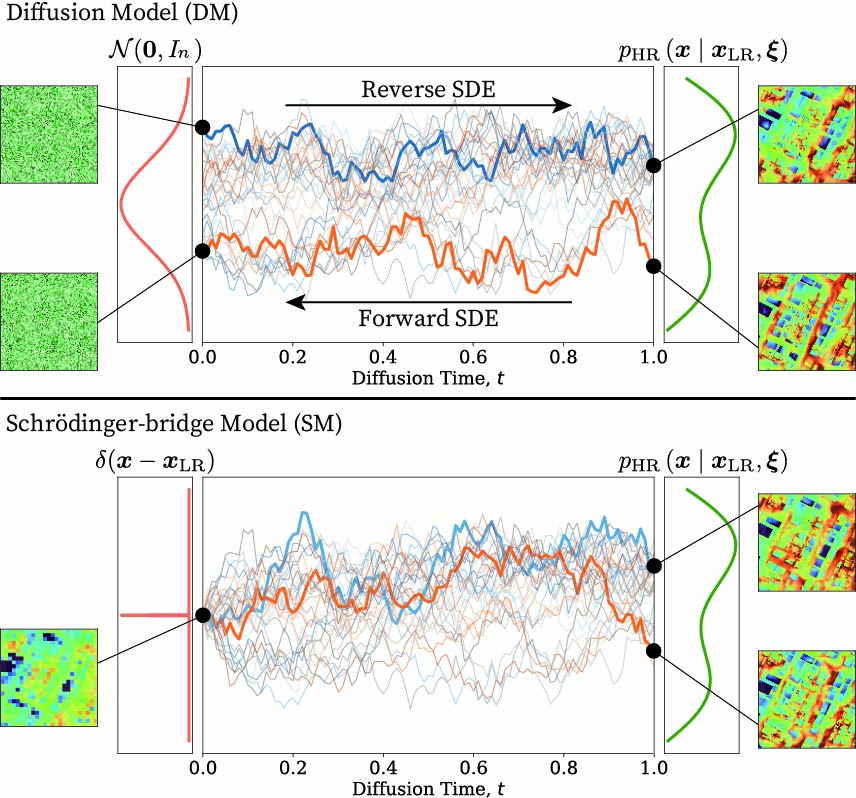}
    \caption{Schematic of data transformation via SDEs. (Top) diffusion model \citep[DM;][]{Ho+2020}; (Bottom) Schrödinger-bridge model \citep[SM;][]{Chen+2024}.}
    \label{fig1}
\end{figure}

We employ an SB-based SR model, simply the Schrödinger-bridge model (SM; see Appendix B for details). This model, originally proposed for forecasting tasks \citep{Chen+2024}, can be adapted to SR. Specifically, we obtain an SDE solution to a particular SB problem that transforms the point mass $\delta(\bm{x}-\bm{x}_\mathrm{LR})$ into $p_\mathrm{HR}(\bm{x}\mid \bm{x}_\mathrm{LR}, \bm{\xi})$ (Fig. \ref{fig1}). This SDE is learned by a neural network, and its numerical integration generates HR samples directly from LR inputs.

For comparison, we use a DM as the baseline \citep{Ho+2020}. This DM transforms a standard Gaussian ${\cal N}(\bm{0}, I_n)$ into $p_\mathrm{HR}(\bm{x}\mid \bm{x}_\mathrm{LR}, \bm{\xi})$ (see Appendix C). Specifically, the DM employs a forward SDE that gradually perturbs the HR data with noise (Fig. \ref{fig1}), and a neural network learns the corresponding reverse dynamics---strictly speaking, the score function. Integrating the learned reverse SDE generates HR samples from noise (Fig. \ref{fig1}). In this formulation, $\bm{x}_\mathrm{LR}$ is input to the neural network together with $\bm{\xi}$ as auxiliary information. Note that the SM requires only a single directional SDE \citep{Chen+2024}, whereas some Schrödinger-bridge formulations require both forward and reverse SDEs \citep[e.g.,][]{Chen+2022}. The behavior of the SM also depends on its reference SDE, and we summarize how this reference differs from that of the DM in Appendix D.

SMs are considered more efficient than DMs for two reasons. First, SMs generate HR samples from $\bm{x}_\mathrm{LR}$, whereas DMs generate them from Gaussian noise. Compared with noise lacking spatial structure, $\bm{x}_\mathrm{LR}$ is expected to have spatial structure similar to that of HR samples. Thus, the transformation with the SM can be performed with fewer steps \citep[e.g.,][]{Liu+2023}. Hereafter, we refer to this step as a diffusion-time step to distinguish it from physical time. Second, DMs do not represent exact SB solutions and, in principle, require many diffusion-time steps \citep{DeBortoli+2021}. Specifically, both the endpoint of the forward process and the starting point of the reverse process must be effectively Gaussian \citep{Ikeda+2025}, which demands strong relaxation to Gaussianity---well resolved only with many steps \citep{DeBortoli+2021}. In contrast, SMs have no such constraint and are expected to attain high accuracy with fewer steps. Further detailed comparisons between DMs and SMs are provided in Appendix D.

\section{Methods}\label{sec:methods}

We super-resolved 2-m temperature from 20-m to 5-m resolution (hereafter LR and HR, respectively). These data were obtained from reproduction experiments of extremely hot days during 2013--2020 \citep{Yasuda+Onishi2025a}, conducted with a physics-based micrometeorological model, the Multi-Scale Simulator for the Geoenvironment \citep[MSSG;][]{Onishi+2012, Takahashi+2013, Sasaki+2016, Matsuda+2018}. The LR and HR results were computed in separate simulations with common initial and boundary conditions. Unlike creating LR data by averaging HR results, this setting makes SR more difficult, since the temporal evolution within the computational domain is simulated separately for LR and HR \citep{Wang+2020NIPS, Wang+2021GMD, Yasuda+Onishi2025a}.

\subsection{Data}\label{subsec:data}

\begin{figure}[ht]
    \centering
    \includegraphics[width=0.75\textwidth]{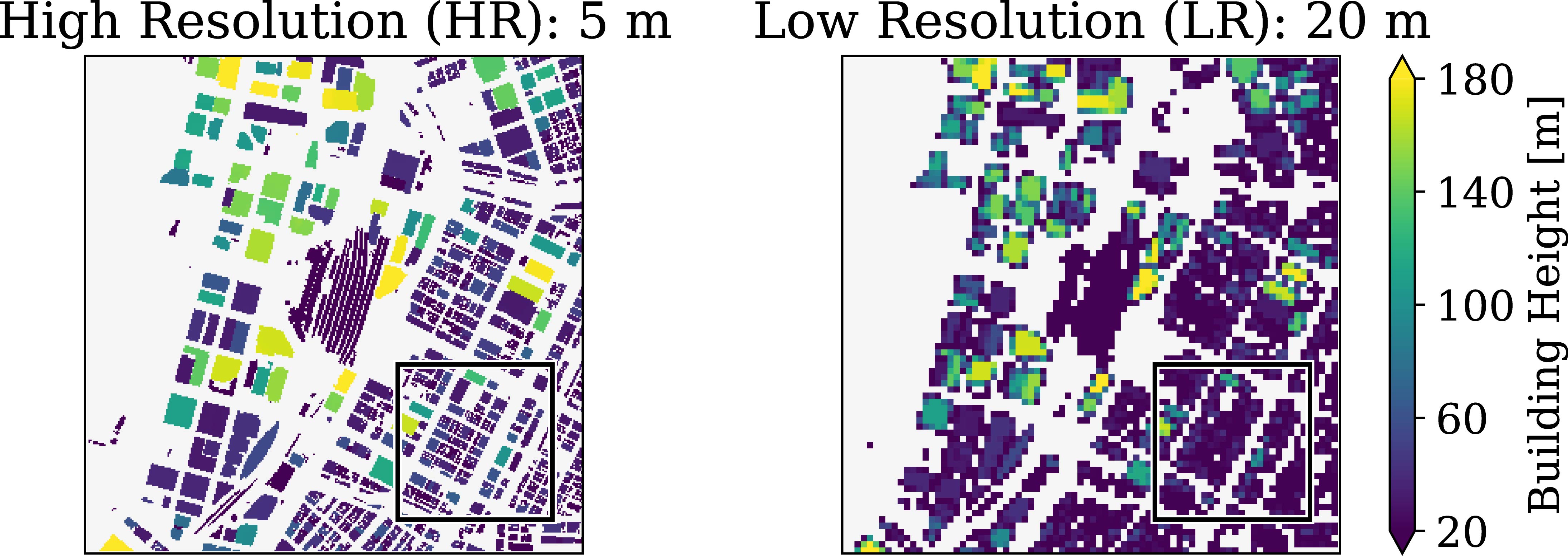}
    \caption{Building height distribution in the computational domain of micrometeorological simulations. (Left) high resolution (HR; 5-m resolution); (Right) low resolution (LR; 20-m resolution). The black squares indicate the area shown in Fig. \ref{fig3}.}
    \label{fig2}
\end{figure}

The datasets cover a 1.6-km square area centered on Tokyo Station (Fig. \ref{fig2}) and consist of MSSG outputs, i.e., 1-min averaged temperature and velocity fields sampled at 1-min intervals. We define one {\it set} (i.e., one dataset) as a pair of LR and HR fields at a single time step. A temporal resolution of 1 min is sufficient to characterize mean flows in street canyons \cite[e.g.,][]{Chew+2018}. For details, see \cite{Yasuda+Onishi2025a}. The LR and HR grids are $80\times 80$ at 20-m resolution and $320\times 320$ at 5-m resolution, respectively. Inputs to the DM and SM consist of LR temperature at 2 m height, together with LR temperature and LR three-dimensional velocity in the lowest seven vertical levels. Additionally, HR building height and HR land-use index are used as static inputs. All inputs are stacked along the channel dimension and upsampled to $320\times 320$ using nearest-neighbor interpolation. We denote the LR 2-m temperature by $\bm{x}_\mathrm{LR}$, and all other inputs collectively by $\bm{\xi}$. The output variable is the HR 2-m temperature. Data from 2013--2018 (2,387 sets) were used for training, data from 2019 (493 sets) for validation, and data from 2020 (540 sets) for testing. Hyperparameters were tuned with the validation data (Section \ref{subsec:training}). All results in Section \ref{sec:results-discussion} are based on the test data. We confirmed that similar results are obtained when using a random data partitioning that ignores temporal order, indicating that the results are not sensitive to the choice of data partitioning.

\subsection{Neural networks}\label{subsec:neural-net}

For the DM, we adopted the Palette model \citep{Saharia+2022}, a U-Net based DM designed for SR tasks. To improve accuracy, we applied residual learning \citep{Kim+2016CVPR, Mardani+2025}: the DM is trained to generate $\bm{x}_\mathrm{HR}-\bm{x}_\mathrm{LR}$, and during testing, $\bm{x}_\mathrm{LR}$ is added to the generated samples. For the SM, we used the same U-Net architecture to parameterize the SDE drift (see Appendix B). We briefly summarize this U-Net below; for details, see our public code (Code Availability).

The U-Net consists of four downsampling and four upsampling blocks. Each block halves or doubles the spatial size of the data through two-dimensional convolutions. In downsampling, the number of channels increases to 32, 64, 128, and 256, whereas in upsampling the channels decrease in reverse order. After downsampling, we applied multi-head self-attention with eight heads \citep{Vaswani+2017}. Diffusion-time steps are encoded using sinusoidal embeddings \citep{Vaswani+2017} and passed to each block as scale and shift parameters \citep{Perez+2018}. We also conducted experiments in which the number of channels or layers in the U-Net was increased and confirmed that the results were not highly sensitive to these changes (details not shown).

Although various acceleration methods have been developed for DMs and SMs \citep[e.g.,][]{Song+2021b, Karras+2022, Boffi+2025, Wang+2025}, we did not use them. For example, implicit acceleration methods (e.g., DDIM) have been proposed for both DMs and SMs \citep{Song+2021b, Wang+2025}. However, to highlight the differences between the DM and SM frameworks, we followed their original formulations \citep{Ho+2020, Chen+2024}, using the same U-Net architecture.

To evaluate inference efficiency, we varied the number of diffusion-time steps $N_T$ from 3 to 1,000 and solved the SDEs with the Euler--Maruyama method \citep{Chen+2024}. For example, since the DM and SM use the same U-Net architecture, the neural networks were evaluated 50 and 10 times when $N_T=50$ for the DM and $N_T=10$ for the SM, respectively. In this setting, the computational cost and runtime of the SM are approximately one-fifth those of the DM.

\subsection{Training}\label{subsec:training}

The DM and SM were trained with AdamW \citep{Loshchilov+Hutter2019} using a learning rate of $1\times 10^{-4}$, a batch size of 32, and 1,000 epochs. The SM loss was the mean squared error (MSE) between the SDE drift and its U-Net approximation, whereas the DM loss was the MSE for denoising score matching (see Appendices B and C).

The following hyperparameters were tuned to minimize the root mean square error (RMSE) on the validation data: number of epochs, learning rate, batch size, and noise amplitudes for both the DM and SM. These parameters were fixed across all experiments after tuning, and the results were not highly sensitive to their exact values. For the SM, the noise amplitude decreased from $2.0\times 10^{-1}$ to $0$ with diffusion time ($\gamma_t$; see Appendix B). For the DM, the drift coefficient decreased linearly from $1\times 10^{1}$ to $1\times 10^{-3}$ ($\lambda_t$; see Appendix C), giving a maximum noise amplitude of $\sqrt{1\times 10^{1}}\approx 3.16$. We confirmed that the tuned values were comparable to those reported in previous studies \citep{Ho+2020, Saharia+2022, Saharia+2023, Chen+2024}.

\subsection{Evaluation metrics}\label{subsec:evaluation-metrics}

SR accuracy was evaluated with two metrics: RMSE, which quantifies pointwise error, and the structural similarity index measure (SSIM) loss, which assesses pattern similarity. For both metrics, smaller values indicate results closer to the ground truth. These metrics are widely used in SR research \citep{Chauhan+2023, Lepcha+2023}. In practice, we generated one HR sample for each ground-truth datum, computed the metrics, and averaged them spatially and temporally over the test data (540 sets). The mean values were nearly independent of the random noise used during sample generation.

Inference statistics were evaluated with 64-member ensembles for each ground-truth datum. Similar results were obtained with 32-member ensembles. For the SM, all members were initialized from the same LR state (2-m temperature), and diversity arose from stochasticity during SDE integration (Fig. \ref{fig1}). For the DM, each member was initialized with different Gaussian noise and further perturbed during integration (Fig. \ref{fig1}).

We report the spread--skill ratio (Spread/RMSE) and the rank histogram, both of which are standard diagnostics \citep{Wilks2011, Fortin+2014}. The spread--skill ratio compares the ensemble spread (i.e., standard deviation) with the RMSE between the ensemble mean and the ground truth. Ratios close to 1 indicate appropriate uncertainty, whereas ratios less than 1 (or greater than 1) indicate underdispersion (or overdispersion). The rank histogram collects the ranks (1--65) of ground-truth values relative to 64 members. Flat histograms indicate reliable dispersion, whereas U-shaped histograms---often observed for deep learning models \citep[e.g.,][]{Mardani+2025}---indicate underdispersion; that is, a tendency for the ground truth to lie outside the ensemble range. While rank histograms are constructed from ranks at each grid point, the spread--skill ratio is computed from RMSEs and spreads after spatial averaging.

\section{Results and discussion}\label{sec:results-discussion}

\subsection{Accuracy for SR inference}

\begin{figure}[ht]
    \centering
    \includegraphics[width=1.2\textwidth]{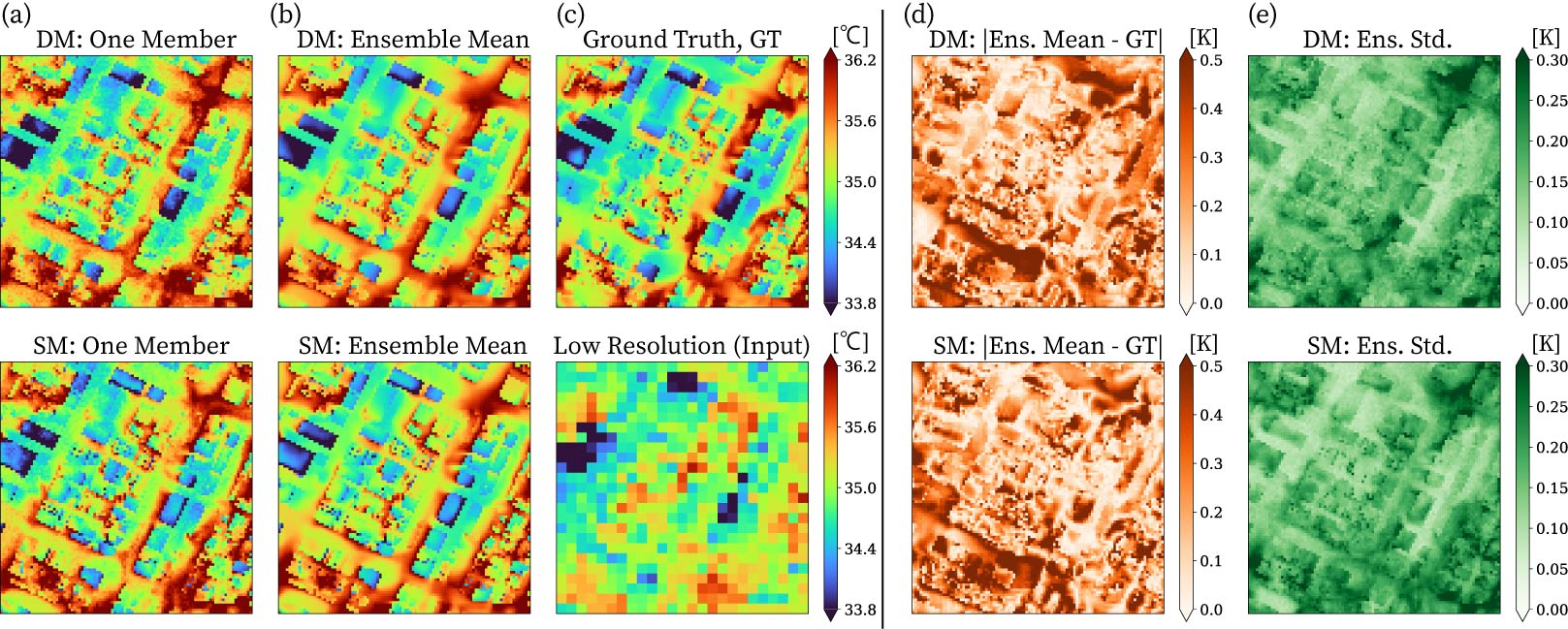}
    \caption{SR for 2-m temperature over a 500 m $\times$ 500 m area in Yaesu, Tokyo (see Fig. \ref{fig2}) at 2020-08-15 14:46+09:00: (a) an ensemble member, (b) ensemble mean, (c) ground-truth datum (i.e., HR datum), (d) absolute error between the ensemble mean and ground truth, and (e) ensemble spread (i.e., standard deviation). The top row shows diffusion model (DM) results, and the bottom row shows Schrödinger-bridge model (SM) results. In the middle column (c), the bottom row shows the input LR 2-m temperature. The DM uses $N_T=50$, whereas the SM uses $N_T=10$. The 2-m temperature is evaluated 2 m above building surfaces, or above the ground in the absence of buildings.}
    \label{fig3}
\end{figure}

Figure \ref{fig3} shows an example of SR. Compared with the input LR field, the HR simulation resolves fine-scale buildings and the associated temperature patterns. Both the DM and SM reproduce these HR patterns well, with $N_T=50$ for the DM and $N_T=10$ for the SM (Figs. \ref{fig3}a--\ref{fig3}c); thus, the SM attains comparable accuracy at approximately one-fifth the computational cost of the DM. A single ensemble member closely resembles the ensemble mean at small scales, and averaging introduces only mild blurring. This likely reflects that HR building geometry governs the spatial scales of temperature \citep{Yasuda+Onishi2025b}, and both the DM and SM use this static HR building field as an important auxiliary input. Comparing the distributions of absolute error and ensemble spread, we find that spreads tend to be larger in regions with larger errors (Figs. \ref{fig3}d and \ref{fig3}e). Indeed, the Pearson correlation between absolute error and spread is about 0.32 for both the DM and SM, which is significantly positive. To further examine this relationship, we also evaluated RMSE and ensemble spread after excluding building areas. When focusing on ground-only regions, the correspondence between RMSE and spread becomes clearer, implying that building-induced variability can partially obscure this relationship.

\begin{figure}[ht]
    \centering
    \includegraphics[width=1.0\textwidth]{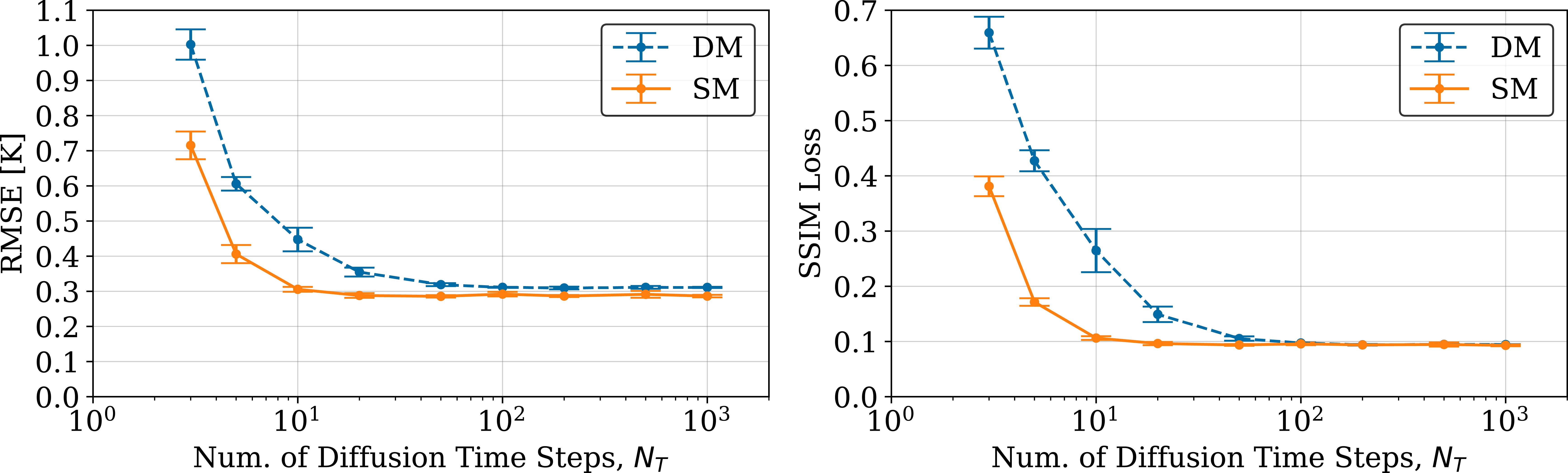}
    \caption{Dependence of mean test errors for 2-m temperature on the number of diffusion-time steps $N_T$. For example, with $N_T=10$, the diffusion time $t \in [0, 1]$ is divided into 10 steps, and the SDE is numerically integrated using the Euler--Maruyama method. Mean test errors were calculated by performing single-member inference for each ground truth and then averaging over all test data (540 sets). Error bars indicate the standard deviations over five experiments with different random initializations of the U-Nets.}
    \label{fig4}
\end{figure}

Figure \ref{fig4} shows the dependence of mean test errors (RMSE and SSIM loss) on the number of diffusion-time steps $N_T$. The DM is sensitive to $N_T$, with mean test errors increasing around $N_T \sim 50$, whereas the SM maintains nearly constant errors down to $N_T \sim 10$. Similar results have been reported in computer vision \citep{Liu+2023}. In principle, DMs require large $N_T$ \citep{DeBortoli+2021}; insufficient $N_T$ can lead to large errors \citep{DeBortoli+2021, Ikeda+2025}. By contrast, SMs have no such limitation and show only weak dependence on $N_T$, except at extremely low values \citep{DeBortoli+2021, Liu+2023}. Our results are consistent with these previous findings.

Table \ref{tab1} reports the mean test errors for the DM ($N_T=50$) and the SM ($N_T=10$). These values correspond to those shown in Fig. \ref{fig4}, with the same $N_T$ settings as in Fig. \ref{fig3}. The SM attains accuracy comparable to that of the DM at about one-fifth the computational cost.

\begin{table}[ht]
    \centering
    \caption{Mean test errors for the DM and SM. Errors were computed by performing single-member inference for each ground truth and then averaging spatially and temporally over all test data (540 sets). Standard deviations were obtained from five experiments with different random initializations of the U-Nets.}
    \label{tab1}
    \begin{tabular}{@{}lll@{}}
    \toprule
                            & RMSE [K]                   & SSIM Loss                  \\
    \midrule
    DM (1 Member, $N_T=50$) &         0.319 $\pm$ 0.004  &        {0.105 $\pm$ 0.004} \\
    SM (1 Member, $N_T=10$) &        {0.306 $\pm$ 0.007} &         0.106 $\pm$ 0.003  \\
    \botrule
    \end{tabular}
\end{table}

\subsection{Variance for ensemble SR inference}

\begin{figure}[ht]
    \centering
    \includegraphics[width=0.75\textwidth]{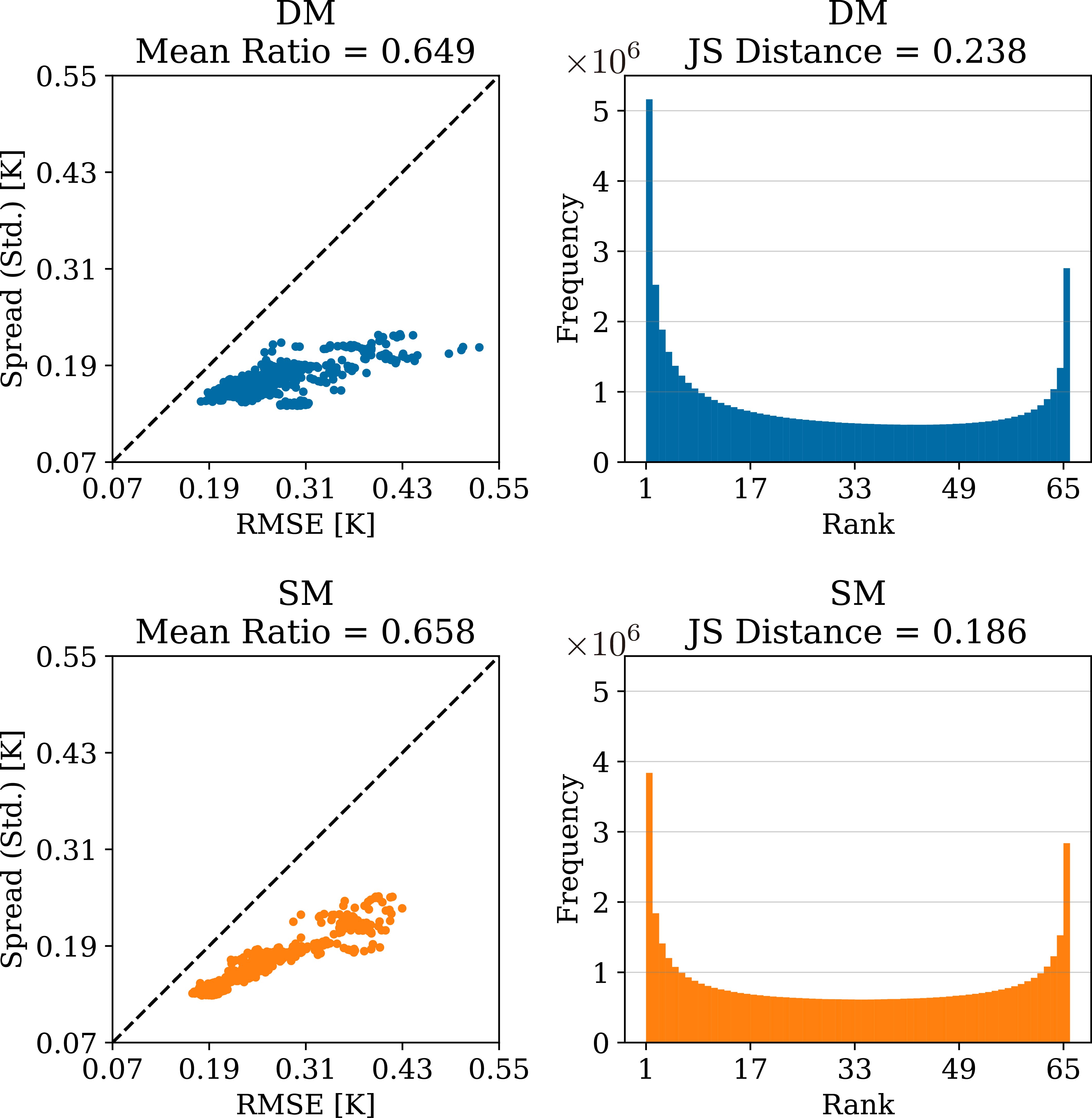}
    \caption{(Left) Scatter plots of ensemble spread versus ensemble mean RMSE; (Right) rank histograms. (Top) The DM with $N_T=50$; (Bottom) the SM with $N_T=10$. The number of ensemble members was 64. In the left column, mean spread--skill ratios (Spread/RMSE) are shown in the figure titles, and in the right column, Jensen--Shannon distances \citep[JS distances;][]{Endres+Schindelin2003} from a uniform distribution are shown. The closer the spread--skill ratio is to 1 and the closer the Jensen--Shannon distance is to 0, the more appropriate the ensemble variance.}
    \label{fig5}
\end{figure}

Figure \ref{fig5} shows scatter plots of RMSE versus spread and rank histograms. Both the DM and SM exhibit spread--skill ratios less than 1 and U-shaped rank histograms, indicating ensemble underdispersion. Such underdispersion is common in deep learning models for meteorological problems \citep[e.g.,][]{Mardani+2025}, motivating the development of models with larger spread. Compared with the DM, the SM is better calibrated: its point cloud lies closer to the diagonal, with a spread--skill ratio (Spread/RMSE) of 0.658, larger than 0.649 for the DM. Rank histograms are also flatter for the SM; the Jensen--Shannon distance \citep{Endres+Schindelin2003} from a perfectly flat uniform distribution is 0.186 for the SM, smaller than 0.238 for the DM.

To assess significance, we used the results from U-Nets initialized with different random weights and computed the means and standard deviations of the spread--skill ratio and the Jensen--Shannon distance (Table \ref{tab2}). The spread--skill ratio for the SM is closer to 1 than that for the DM, and the Jensen--Shannon distance for the SM is closer to 0. We also computed another common diagnostic, the continuous ranked probability score \citep[CRPS;][]{Hersbach2000, Wilks2011}. The CRPS is smaller for the SM, consistent with the other diagnostics (Table \ref{tab2}). Thus, ensemble inference with the SM yields more appropriate statistics, exhibiting milder underdispersion than the DM.

\begin{table}[ht]
    \centering
    \caption{Spread--skill ratios (Spread/RMSE), Jensen--Shannon distances, and continuous ranked probability scores (CRPS) for the DM and SM. These quantities were computed by performing 64-member inference for each ground truth and then averaging spatially and temporally over the test data (540 sets). Standard deviations were obtained as in Table \ref{tab1}.}
    \label{tab2}
    \begin{tabular}{@{}llll@{}}
    \toprule
                              & Spread--Skill Ratio              & Jensen--Shannon Distance   & CRPS [K]                  \\
    \midrule
    DM (64 Members, $N_T=50$) &         0.641 $\pm$ 0.024        &         0.234 $\pm$ 0.025  &         0.151 $\pm$ 0.004 \\
    SM (64 Members, $N_T=10$) & \textbf{0.657 $\pm$ 0.013}       & \textbf{0.189 $\pm$ 0.015} & \textbf{0.140 $\pm$ 0.003}\\
    \botrule
    \end{tabular}
\end{table}

Theoretically, both the DM and SM approximate the conditional distribution $p_\mathrm{HR}(\bm{x}\mid\bm{x}_\mathrm{LR}, \bm{\xi})$ \citep{Song+2021a, Chen+2024}. Indeed, when using the same U-Net architecture, they achieve comparable accuracy (Fig. \ref{fig4} and Table \ref{tab1}). With respect to ensemble statistics, although both the DM and SM exhibit underdispersion, the SM shows a wider spread, indicating that its ensemble statistics are superior (Fig. \ref{fig5} and Table \ref{tab2}). When finite data are used, transformation from LR inputs may be easier to learn than transformation from noise. This ease of learning for the SM is considered to account for the difference in ensemble statistics.

\subsection{Total inference time}\label{subsec:inference-time}

Finally, we discuss the total inference time for 60-min predictions. We first report the wall-clock times for physics-based MSSG simulations, measured on the Earth Simulator at the Japan Agency for Marine-Earth Science and Technology (JAMSTEC), equipped with AMD EPYC 7742 CPUs. On average, HR simulations required 206 min with 256 CPU cores, whereas LR simulations required 6.19 min with 40 CPU cores \citep{Yasuda+Onishi2025a}. We then measured the average SM inference time on a local workstation equipped with an NVIDIA L40S GPU. For 60-min LR data (60 sets), single-member inference took 7.29 s. Thus, the hybrid method combining LR simulation and SM inference completed 60-min predictions of 2-m temperature in 6.31 min, reducing HR computation time to 3.06\% (a 32.6-fold speedup). This factor is comparable to values reported for recent surrogate models in urban airflow simulations \citep{Shao+2023BAE, Peng+2024BAE}.

For multi-member ensemble inference, technical considerations are required. With 64 members, inference took 7.80 s per LR dataset. To process this computation in real time, SM inference must be performed as soon as LR data become available, i.e., in parallel with the LR simulation by MSSG. This would enable completion of 60-min predictions in about 8 min ($\sim$7.80 min). These estimates suggest the feasibility of real-time ensemble micrometeorological prediction using the SM.

\section{Conclusions}\label{sec:conclusions}

We applied a Schrödinger-bridge model \citep[SM;][]{Chen+2024} to super-resolve urban 2-m temperature and showed that the SM achieves accuracy comparable to that of a diffusion model \citep[DM;][]{Ho+2020}, while requiring only about one-fifth the computational cost. We further showed that SM ensembles yield larger spreads and more appropriate statistics than DM ensembles.

Previous studies demonstrate that DMs can handle other variables, such as precipitation and wind velocity \cite[e.g.,][]{Hess+2025, Mardani+2025}, with variable-specific preprocessing where needed \cite[e.g., log transformation for precipitation;][]{Ling+2024}. These findings suggest that Schrödinger-bridge techniques may be extended to other meteorological variables under similar considerations. The present study did not incorporate acceleration methods in order to focus on the original formulations of the SM and DM. Recent results \cite[e.g.,][]{Wang+2025} suggest that SMs may remain more efficient than DMs even when both are accelerated. Verifying the impact of such acceleration in meteorological applications is an important direction for future work.

\backmatter




\bmhead{Acknowledgements}

The micrometeorological simulations and deep learning experiments were performed on the Earth Simulator at the Japan Agency for Marine-Earth Science and Technology (JAMSTEC). Additional deep learning experiments were conducted on a local workstation equipped with an NVIDIA L40S GPU.

\section*{Declarations}

\subsection*{Funding}

This work was supported by the JSPS KAKENHI (grant number 25H00715).

\subsection*{Competing interests}

The authors declare no competing financial interests or personal relationships that could have influenced the work reported in this paper.

\subsection*{Data availability}

The data that support the findings of this study are available from the corresponding author upon reasonable request.

\subsection*{Code availability}\label{subsec:code-availability}

The source code for the deep learning models is preserved in the Zenodo repository (\url{https://doi.org/10.5281/zenodo.17346029}) and is openly developed in the GitHub repository (\url{https://github.com/YukiYasuda2718/schrodinger-bridge-sr-micrometeorology/releases/tag/v0.1.0}).


\begin{appendices}

\section{List of abbreviations}

Table S1 summarizes the abbreviations used in the manuscript, along with their full terms, listed in alphabetical order.

\begin{table}[ht]
    \centering
    \caption{List of abbreviations.}
    \renewcommand{\thetable}{A\arabic{table}}\setcounter{table}{1}
    \begin{tabular}{@{}ll@{}}
    \toprule
    Abbreviation & Full Term          \\
    \midrule
    CPU & Central Processing Unit \\
    CRPS & Continuous Ranked Probability Score \\
    DDIM & Denoising Diffusion Implicit Model \\
    DM & Diffusion Model \\
    GPU & Graphics Processing Unit \\
    HR & High Resolution \\
    JS & Jensen--Shannon \\
    LR & Low Resolution \\
    MSE & Mean Squared Error \\
    MSSG & Multi-Scale Simulator for the Geoenvironment \\
    RMSE & Root Mean Square Error \\
    SB & Schrödinger Bridge \\
    SDE & Stochastic Differential Equation \\
    SI & Supplementary Information \\
    SM & Schrödinger-bridge Model \\
    SR & Super-Resolution \\
    SSIM & Structural Similarity Index Measure \\
    \botrule
    \end{tabular}
\end{table}

\section{Schrödinger-bridge model (SM)}

The Schrödinger-bridge model \citep[SM;][]{Chen+2024} directly transforms $\bm{x}_\mathrm{LR}$ into HR samples using an SDE. This SDE corresponds to entropy-regularized optimal transport \citep{Friesecke2024} from the point mass $\delta(\bm{x}-\bm{x}_\mathrm{LR})$ to $p_\mathrm{HR}(\bm{x}\mid\bm{x}_\mathrm{LR},\bm{\xi})$, thereby solving a particular SB problem \citep{Chen+2024}. Equivalently, among SDEs that transform $\delta(\bm{x}-\bm{x}_\mathrm{LR})$ to $p_\mathrm{HR}(\bm{x}\mid\bm{x}_\mathrm{LR},\bm{\xi})$, we consider the most efficient one, which is defined as the SDE process closest to a reference process. This reference process is discussed in Appendix C.

The SDE for the SM \citep{Chen+2024} is given by
\begin{equation}
    d \bm{x}_t = \left[ \bm{b}(\bm{x}_t, \bm{x}_\mathrm{LR}, \bm{\xi}) +\frac{1}{2}(g_t^2 - \gamma_t^2) \nabla_{\bm{x}_t} \ln p(\bm{x}_t|\bm{x}_\mathrm{LR}, \bm{\xi}) \right] dt + g_t d\bm{W}_t, \label{eq:sde-sm}
\end{equation}
where the diffusion time is $t \in [0,1]$, the state vector is $\bm{x}_t \in \mathbb{R}^n$, and $\bm{W}_t \in \mathbb{R}^n$ denotes a Wiener process. The scalars $g_t$ and $\gamma_t$ are defined as
\begin{align}
    g_t &= \epsilon \sqrt{(3-t)(1-t)}, \\
    \gamma_t &= \epsilon (1 - t).
\end{align}
The score function $\nabla_{\bm{x}_t}\ln p(\bm{x}_t \mid \bm{x}_\mathrm{LR}, \bm{\xi})$ is computed explicitly from $\bm{b}(\bm{x}_t, \bm{x}_\mathrm{LR}, \bm{\xi})$ \citep{Chen+2024}. Thus, given the drift term $\bm{b}$ in Eq. (\ref{eq:sde-sm}), HR samples $\bm{x}_{t=1}$ can be generated from $\bm{x}_{t=0}$ ($=\bm{x}_\mathrm{LR}$) by solving Eq. (\ref{eq:sde-sm}). Theoretically, these HR samples follow the conditional distribution $p_\mathrm{HR}(\bm{x} \mid \bm{x}_\mathrm{LR}, \bm{\xi})$.

The function $\bm{b}(\bm{x}_t, \bm{x}_\mathrm{LR}, \bm{\xi})$ is represented by a neural network $\hat{\bm{b}}$ and estimated through supervised learning with the following loss function $\mathcal{L}_\mathrm{SM}$ \citep{Chen+2024}:
\begin{equation}
    \mathcal{L}_\mathrm{SM} = \mathbb{E}\left[ \left\lVert \left(\dot{\alpha}_t \bm{x}_\mathrm{LR} + \dot{\beta}_t \bm{x}_\mathrm{HR} + \dot{\gamma}_t \bm{W}_t\right) - \hat{\bm{b}}(\bm{I}_t, \bm{x}_\mathrm{LR}, \bm{\xi}) \right\rVert^2 \right],
\end{equation}
where the expectation is taken over $\bm{x}_\mathrm{LR}$, $\bm{x}_\mathrm{HR}$, $\bm{W}_t$, and $t$. Dots (e.g., $\dot{\alpha}_t$) denote derivatives with respect to $t$. The functions $\alpha_t, \beta_t \in \mathbb{R}$ and $\bm{I}_t \in \mathbb{R}^n$ are defined as
\begin{align}
    \alpha_t &= 1 - t, \\
    \beta_t &= t^2, \\
    \bm{I}_t &= {\alpha}_t \bm{x}_\mathrm{LR} + {\beta}_t \bm{x}_\mathrm{HR} + {\gamma}_t \bm{W}_t,
\end{align}
where $\bm{I}_t$ is called a stochastic interpolant \citep{Albergo+2023, Albergo+2024}. This quantity represents an interpolation point ($\alpha_t \bm{x}_\mathrm{LR} + \beta_t \bm{x}_\mathrm{HR}$) between LR--HR data pairs with added noise $\gamma_t \bm{W}_t$.

Intuitively, the trained $\hat{\bm{b}}(\bm{x}_t, \bm{x}_\mathrm{LR}, \bm{\xi})$ represents the velocity pointing from the current state $\bm{x}_t$ toward the HR data. In this sense of learning the velocity of data transformation, the model of \cite{Chen+2024} resembles flow matching \citep{Lipman+2023}. More generally, the stochastic interpolant framework \citep{Albergo+2023, Albergo+2024} used in \cite{Chen+2024} encompasses both flow matching and diffusion models.

\section{Diffusion model (DM)}

We summarize the formulation of denoising diffusion probabilistic models (simply, diffusion models; DMs) \citep{Sohl-Dickstein+2015, Ho+2020}. For comparison with the SM, we use the continuous diffusion-time formulation \citep{Song+2021a}. The DM transforms standard Gaussian noise into HR samples using forward and reverse SDEs:
\begin{align}
    d \bm{y}_t &= -\frac{1}{2} \lambda_t \bm{y}_t \,dt + \sqrt{\lambda_t} \, d \bm{W}_t, \label{eq:sde-dm-forward} \\
    d \bm{y}_t &= \left[-\frac{1}{2} \lambda_t \bm{y}_t - \lambda_t \nabla_{\bm{y}_t} \ln p_t(\bm{y}_t | \bm{x}_\mathrm{LR}, \bm{\xi}) \right] dt + \sqrt{\lambda_t} \, d \bm{W}_t, \label{eq:sde-dm-backward}
\end{align}
where the diffusion time is $t\in[0,T]$, the state vector is $\bm{y}_t\in\mathbb{R}^n$, $\bm{W}_t$ denotes a Wiener process, and $\lambda_t \in \mathbb{R}$ is a prescribed increasing function of $t$. Since this study uses residual learning \citep{Kim+2016CVPR, Mardani+2025}, $\bm{y}_{t=0}$ denotes the residual $\bm{x}_\mathrm{HR}-\bm{x}_\mathrm{LR}$. The endpoint $\bm{y}_{t=T}$ corresponds to Gaussian noise ($\bm{y}_{t=T} \sim {\cal N}(0, I_n)$), where $I_n$ is the $n$-dimensional identity matrix.

The forward SDE (\ref{eq:sde-dm-forward}) maps the residual $\bm{x}_\mathrm{HR}-\bm{x}_\mathrm{LR}$ to noise, whereas the reverse SDE (\ref{eq:sde-dm-backward}) maps noise to the residual. The reverse SDE is analytically derived from the forward SDE and includes the score function $\nabla_{\bm{y}_t} \ln p_t(\bm{y}_t | \bm{x}_\mathrm{LR}, \bm{\xi})$ \citep{Anderson1982, Hirono+2024}. If the score function is known, HR samples can be generated from noise by solving the reverse SDE. Theoretically, this HR sample follows the conditional distribution $p_\mathrm{HR}(\bm{x}\mid\bm{x}_\mathrm{LR},\bm{\xi})$.

The score function is learned via denoising score matching \citep{Vincent2011, Ho+2020}. The forward SDE has the analytic solution
\begin{equation}
    \bm{y}_t = \mu_t\bm{y}_0 + \sigma_t \bm{\eta},    
\end{equation}
where $\mu_t$ and $\sigma_t$ are obtained explicitly from $\lambda_t$ \citep[e.g.,][]{Gardiner2009}, and $\bm{\eta} \in \mathbb{R}^n$ represents Gaussian noise ($\bm{\eta}\sim{\cal N}(0,I_n)$). The loss function $\mathcal{L}_\mathrm{DM}$ for denoising score matching is given by \citep{Ho+2020}
\begin{equation}
    \mathcal{L}_\mathrm{DM} = \mathbb{E} \left[ \left\lVert -\frac{\bm{\eta}}{\sigma_t} - \hat{\bm{s}}_t(\bm{y}_t,\bm{x}_\mathrm{LR}, \bm{\xi}) \right\rVert^2 \right],
\end{equation}
where the expectation is taken over $t$ and $\bm{\eta}$. The score function is approximated with a neural network, $\hat{\bm{s}}_t(\bm{y}_t,\bm{x}_\mathrm{LR}, \bm{\xi}) \approx \nabla_{\bm{y}_t} \ln p_t(\bm{y}_t | \bm{x}_\mathrm{LR}, \bm{\xi})$. The trained score function $\hat{\bm{s}}$ is approximately proportional to noise $\bm{\eta}$. Thus, the transformation by the reverse SDE (\ref{eq:sde-dm-backward}) is sometimes explained as {\it denoising} by referring to the drift term $\left[-0.5 \lambda_t \bm{y}_t - \lambda_t \nabla_{\bm{y}_t} \ln p_t(\bm{y}_t | \bm{x}_\mathrm{LR}, \bm{\xi}) \right] dt$. However, this explanation is not completely accurate because the reverse SDE also includes the addition of noise $\sqrt{\lambda_t} d \bm{W}_t$ \citep{Ho+2020, Song+2021a}. Mathematically, the reverse SDE exists first, and the score function is learned through denoising score matching \citep{Song+2021a, Hirono+2024}. This learning corresponds to estimating the noise $\bm{\eta}$ by minimizing the loss function $\mathcal{L}_\mathrm{DM}$.

\section{Detailed Comparison of SM and DM}

The SM and DM generate HR samples via their respective SDEs, but there are three key differences. First, the SM transforms the point mass $\delta(\bm{x}-\bm{x}_\mathrm{LR})$ into $p_\mathrm{HR}(\bm{x}\mid\bm{x}_\mathrm{LR},\bm{\xi})$ \citep{Chen+2024}, whereas the DM transforms ${\cal N}(0,I_n)$ into the same conditional distribution \citep{Song+2021a}. Figure 1 in the main text schematically illustrates these transformations. Compared with Gaussian noise lacking spatial structure, $\bm{x}_\mathrm{LR}$ is expected to have spatial structure closer to that of HR samples. Owing to this difference in initial conditions, the SM is expected to generate data more efficiently \citep[e.g.,][]{Liu+2023}.

Second, unlike the SM, the DM provides only an approximate solution to a Schrödinger bridge (SB) problem \citep{DeBortoli+2021, Chen+2022}. In the forward SDE (\ref{eq:sde-dm-forward}), the final distribution converges to ${\cal N}(0,I_n)$ in the limit of infinite diffusion time ($T \rightarrow \infty$). In this limit, the DM becomes an exact solution to the SB. Since $T$ is finite in practice, however, the DM represents only an approximate solution to the SB. This approximation, in principle, necessitates a large number of diffusion-time steps \citep{DeBortoli+2021}, leading to inefficient and time-consuming inference.

The third difference lies in the reference process. SB problems derive optimal transformations in the sense of being closest to reference processes \citep{Leonard2014}. Both the DM and SM describe the reference processes as SDEs with linear drifts, where the final distributions converge to ${\cal N}(0,I_n)$ \citep{DeBortoli+2021, Chen+2024}. However, the SM explicitly incorporates $\bm{x}_\mathrm{LR}$ in the reference SDE, making the SB problem {\it conditional} on $\bm{x}_\mathrm{LR}$. In contrast, the DM does not use such conditioning. Conditional methods are particularly effective for large data, such as images \citep{Liu+2023, Chen+2024, Wang+2025}, and allow more efficient learning and inference compared with unconditional SB problems \citep{DeBortoli+2021, Chen+2022}. Finally, we note that the conditional reference process used here follows the specific construction of \cite{Chen+2024} and represents only one possible choice. There is currently no general guideline for selecting reference processes, and alternative constructions \cite[e.g.,][]{Liu+2023} may lead to different properties or efficiencies.

\end{appendices}


\bibliography{references}


\end{document}